\documentclass[dvips]{acta}
\usepackage{supertabular,lscape,epsfig}
\usepackage{amssymb}
\usepackage{amsmath}
\usepackage{float}
\mathchardef\mhyphen="2D
\usepackage{placeins}
\DeclareSymbolFont{ppa}{OT1}{ppl}{m}{it}
\DeclareMathSymbol{\vv}{\mathalpha}{ppa}{'166}

\SetPages{297}{316}

\SetVol{67}{2017}

\begin{document}
\newcommand\pvalue{\mathop{p\mhyphen {\rm value}}}
\newcommand{\TabApp}[2]{\begin{center}\parbox[t]{#1}{\centerline{
  {\bf Appendix}}
  \vskip2mm
  \centerline{\small {\spaceskip 2pt plus 1pt minus 1pt T a b l e}
  \refstepcounter{table}\thetable}
  \vskip2mm
  \centerline{\footnotesize #2}}
  \vskip3mm
\end{center}}

\newcommand{\TabCapp}[2]{\begin{center}\parbox[t]{#1}{\centerline{
  \small {\spaceskip 2pt plus 1pt minus 1pt T a b l e}
  \refstepcounter{table}\thetable}
  \vskip2mm
  \centerline{\footnotesize #2}}
  \vskip3mm
\end{center}}

\newcommand{\TTabCap}[3]{\begin{center}\parbox[t]{#1}{\centerline{
  \small {\spaceskip 2pt plus 1pt minus 1pt T a b l e}
  \refstepcounter{table}\thetable}
  \vskip2mm
  \centerline{\footnotesize #2}
  \centerline{\footnotesize #3}}
  \vskip1mm
\end{center}}

\newcommand{\MakeTableApp}[4]{\begin{table}[p]\TabApp{#2}{#3}
  \begin{center} \TableFont \begin{tabular}{#1} #4 
  \end{tabular}\end{center}\end{table}}

\newcommand{\MakeTableSepp}[4]{\begin{table}[p]\TabCapp{#2}{#3}
  \begin{center} \TableFont \begin{tabular}{#1} #4 
  \end{tabular}\end{center}\end{table}}

\newcommand{\MakeTableee}[4]{\begin{table}[htb]\TabCapp{#2}{#3}
  \begin{center} \TableFont \begin{tabular}{#1} #4
  \end{tabular}\end{center}\end{table}}

\newcommand{\MakeTablee}[5]{\begin{table}[htb]\TTabCap{#2}{#3}{#4}
  \begin{center} \TableFont \begin{tabular}{#1} #5 
  \end{tabular}\end{center}\end{table}}

\newcommand{\MakeTableH}[4]{\begin{table}[H]\TabCap{#2}{#3}
  \begin{center} \TableFont \begin{tabular}{#1} #4 
  \end{tabular}\end{center}\end{table}}

\newcommand{\MakeTableHH}[4]{\begin{table}[H]\TabCapp{#2}{#3}
  \begin{center} \TableFont \begin{tabular}{#1} #4 
  \end{tabular}\end{center}\end{table}}

\newfont{\bb}{ptmbi8t at 12pt}
\newfont{\bbb}{cmbxti10}
\newfont{\bbbb}{cmbxti10 at 9pt}
\newcommand{\uprule}{\rule{0pt}{2.5ex}}
\newcommand{\douprule}{\rule[-2ex]{0pt}{4.5ex}}
\newcommand{\dorule}{\rule[-2ex]{0pt}{2ex}}
\def\thefootnote{\fnsymbol{footnote}}
\begin{Titlepage}
\Title{The OGLE Collection of Variable Stars.\\
Classical, Type II, and Anomalous Cepheids\\
toward the Galactic Center\footnote{Based on observations
obtained with the 1.3-m Warsaw telescope at the Las Campanas Observatory of the Carnegie Institution for Science.}}
\Author{I.~~S~o~s~z~y~ñ~s~k~i$^1$,~~
A.~~U~d~a~l~s~k~i$^1$,~~
M.\,K.~~S~z~y~m~a~ñ~s~k~i$^1$,~~
£.~~W~y~r~z~y~k~o~w~s~k~i$^1$,\\
K.~~U~l~a~c~z~y~k$^2$,~~
R.~~P~o~l~e~s~k~i$^3$,~~
P.~~P~i~e~t~r~u~k~o~w~i~c~z$^1$,~~
S.~~K~o~z~³~o~w~s~k~i$^1$,\\
D.\,M.~~S~k~o~w~r~o~n$^1$,~~
J.~~S~k~o~w~r~o~n$^1$,~~
P.~~M~r~ó~z$^1$,~~
M.~~P~a~w~l~a~k$^1$,~~
K.~~R~y~b~i~c~k~i$^1$\\
and~~A.~~J~a~c~y~s~z~y~n~-~D~o~b~r~z~e~n~i~e~c~k~a$^1$
}
{$^1$Warsaw University Observatory, Al.~Ujazdowskie~4, 00-478~Warszawa, Poland\\
e-mail: soszynsk@astrouw.edu.pl\\
$^2$Department of Physics, University of Warwick, Gibbet Hill Road, Coventry, CV4~7AL,~UK\\
$^3$Department of Astronomy, Ohio State University, 140 W. 18th Ave., Columbus, OH~43210, USA}
\Received{December 5, 2017}
\end{Titlepage}

\Abstract{We present a collection of classical, type~II, and anomalous
  Cepheids detected in the OGLE fields toward the Galactic center. The
  sample contains 87 classical Cepheids pulsating in one, two or three
  radial modes, 924 type~II Cepheids divided into BL~Her, W~Vir, peculiar
  W~Vir, and RV~Tau stars, and 20 anomalous Cepheids -- first such objects
  found in the Galactic bulge. Additionally, we upgrade the OGLE Collection
  of RR~Lyr stars in the Galactic bulge by adding 828 newly identified
  variables. For all Cepheids and RR~Lyr stars, we publish time-series {\it
    VI} photometry obtained during the OGLE-IV project, from 2010 through
  2017.

  We discuss basic properties of our classical pulsators: their spatial
  distribution, light curve morphology, period--luminosity relations, and
  position in the Petersen diagram. We present the most interesting
  individual objects in our collection: a type~II Cepheid with additional
  eclipsing modulation, W~Vir stars with the period doubling effect and the
  RVb phenomenon, a mode-switching RR~Lyr star, and a triple-mode anomalous
  RRd star.}{Stars: variables: Cepheids -- Stars: variables: RR~Lyrae --
  Stars: oscillations -- Galaxy: center -- Catalogs}

\Section{Introduction}
Cepheids and RR~Lyr stars, sometimes collectively called classical
pulsators, undergo radial oscillations driven by the $\kappa$-mechanism in
helium ionization zones. Cepheid variables can be divided into several
subclasses which exhibit markedly different masses, ages, and evolutionary
histories. The youngest ones are classical (or type~I) Cepheids which play
a fundamental role in the calibration of the extragalactic distance scale
thanks to their period--luminosity (PL) relations. On the contrary, type~II
Cepheids belong to an old stellar population, but the exact stage of their
evolution depends on their pulsation periods. BL~Her stars, with periods
ranging between 1~d and 5~d, evolve away from the horizontal branch toward
the asymptotic giant branch. W~Vir stars (periods from 5~d to 20~d) likely
undergo blueward loops from the asymptotic giant branch due to helium shell
flashes. RV~Tau stars (periods longer than 20~d) evolve away from the
asymptotic giant branch toward a white dwarf domain. In turn, the
evolutionary status of anomalous Cepheids is under debate. The two most
popular scenarios are the evolution of a single, intermediate-age,
metal-poor star with mass of 1--2~\MS, and the evolution of coalescent
binary systems of old, low-mass stars.

Classical pulsating stars in the central regions of the Milky Way have been
the subject of extensive research in recent years. On the one hand, the
Optical Gravitational Lensing Experiment (OGLE) published a large catalog
of Cepheids and RR~Lyr stars in the Galactic bulge (Soszyñski \etal 2011,
2013, 2014). On the other hand, near-infrared surveys, like the VISTA
Variables in the V{\'{\i}}a L{\'a}ctea (VVV, Minniti \etal 2010) or
IRSF/SIRIUS (Nagashima \etal 1999), led to the discovery of classical
pulsators in the highly-obscured regions of the bulge, mostly within
$\approx1\arcd$ from the Galactic plane (\eg D{\'e}k{\'a}ny \etal 2015,
Matsunaga \etal 2011, 2013, 2015, 2016).

The previous version of the OGLE catalog (Soszyñski \etal 2011, 2013)
consisted of 32 candidates for classical Cepheids and 357 type~II Cepheids
detected in about 69~square degrees in the Galactic bulge covered by the
OGLE-II and OGLE-III fields. These samples have been successfully used in
several interesting studies on the stellar pulsations itself and on the
structure of our Galaxy. Smolec \etal (2012) reported the discovery of the
first BL~Her stars exhibiting a period-doubling effect -- a phenomenon
theoretically predicted by Buchler and Moskalik (1992) and noticed for the
first time in the OGLE catalog of type~II Cepheids. Feast \etal (2014) used
five fundamental-mode classical Cepheids from the OGLE-III catalog to
discover flared outer disk of the Milky Way. Kovtyukh \etal (2016) studied
metallicity of double-mode classical Cepheids from the OGLE Collection of
Variable Stars (OCVS). Recently, Bhardwaj \etal (2017) matched the OGLE
type~II Cepheids with the VVV near-infrared photometry to determine the
distance to the Galactic center and to investigate the spatial distribution
of the old stellar population in the bulge.

In this paper, we extend the OGLE Collection of Cepheids in the Galactic
bulge by objects identified in the OGLE-IV fields. The new version of our
collection consists of classical Cepheids, type~II Cepheids and, for the
first time, anomalous Cepheids in the central regions of the Milky Way. We
also supplement the OGLE catalog of RR~Lyr stars with 828 newly detected
variables of this type and update the time-series photometry of the
previously published stars.

The rest of the paper is organized as follows. In Section~2, we present the
OGLE photometric data used in this study. Methods applied for the variable
star identification and classification are introduced in
Section~3. Section~4 describes the Cepheid collection itself. In Section~5,
we discuss some interesting features of the published samples of pulsating
stars. Finally, conclusions are presented in Section~6.

\vspace*{-7pt}
\Section{Observations and Data Reduction}
\vspace*{-5pt} 
The OGLE-IV data set used in this investigation was obtained between
March 2010 and August 2017 with the 1.3-m Warsaw telescope located at
Las Campanas Observatory, Chile (operated by the Carnegie Institution
for Science).  A mosaic camera composed of 32 CCD chips with
Johnson-Cousins {\it VI} filters was used, providing a field of view
of 1.4~square degrees with a pixel scale of 0\zdot\arcs26. Details of
the instrument can be found in Udalski \etal (2015).

In total, 121 OGLE-IV fields toward the Galactic center were searched for
classical pulsating stars. Together with the OGLE-II and OGLE-III fields
analyzed by Soszyñski \etal (2011), we studied an area of about 182 square
degrees. Owing to the OGLE observing strategy optimized to detect and
monitor gravitational microlensing events, the number of collected data
points varies significantly from field to field -- from about 40 to 15\,000
epochs per star in the {\it I} passband (median value: 773) and from 0 to
175 in the {\it V}-band (median: 37). Some of these light curves may be
supplemented with observations obtained during the previous phases of the
OGLE project and published by Soszyñski \etal (2011). One should be aware
that small offsets between individual light curves obtained during
different stages of the OGLE project are possible, mostly because of
contamination by neighboring stars in the dense bulge region. This should
be taken into account when merging the photometry from different phases of
the project.

Data reduction was carried out using the Difference Image Analysis software
(Alard and Lupton 1998, Wo¼niak 2000). Detailed descriptions of the
photometric reductions and astrometric calibrations of the OGLE-IV data are
provided by Udalski \etal (2015).

\vspace*{-4pt}
\Section{Selection and Classification of Cepheids}
Each of the 400 million {\it I}-band light curves collected by OGLE in the
Galactic bulge was subjected to a period search procedure based on the
Fourier technique implemented by Z.~Ko³aczkowski in the {\sc Fnpeaks}
code\footnote{\it
  http://helas.astro.uni.wroc.pl/deliverables.php?lang=en\&active=fnpeaks}.
The probed frequency space ranged from 0 to 24 cycles per day, with a step
of 0.00005 cycles per day. For each star the primary period was subtracted
from the original light curve and the period search procedure was repeated
on the residual data. Both periods (primary and secondary) were recorded
with their amplitudes and signal-to-noise ratios. The light curves with the
primary periods between 0.2~d and 100~d and amplitudes larger than 0.05~mag
were fitted with a Fourier cosine series and the low-order Fourier
coefficients $R_{21}$, $\phi_{21}$, $R_{31}$, $\phi_{31}$ (Simon and Lee
1981) were derived.

This dataset became a basis for our variable star detection and
classification procedure. We visually examined the light curves with the
Fourier parameters characteristic for classical pulsators. The final
decision about the classification of each star was made after careful
analysis of the light curve shape quantified by the Fourier
coefficients. In ambiguous cases, we also considered other available
parameters of the stars, like colors and period ratios (for multi-mode
pulsators). Additionally, some candidates for pulsating stars were
identified during the search for eclipsing binary systems in the Milky Way
bulge area (Soszyñski \etal 2016a).

Contrary to the fundamental-mode pulsators, the overtone classical Cepheids
and $\delta$~Sct stars (both single- and multi-mode) constitute continuity
and it is a matter of convention which period will be adopted as a
borderline between both types of variables. In this study, we adopted the
same discrimination period as in other parts of the OCVS: pulsators with
the first-overtone periods shorter than 0.23~d were classified as
$\delta$~Sct stars and will be published elsewhere. The longer-period
(Population~I) pulsators were classified as classical Cepheids.

\begin{figure}[htb]
\includegraphics[width=12.5cm, bb=30 195 565 745]{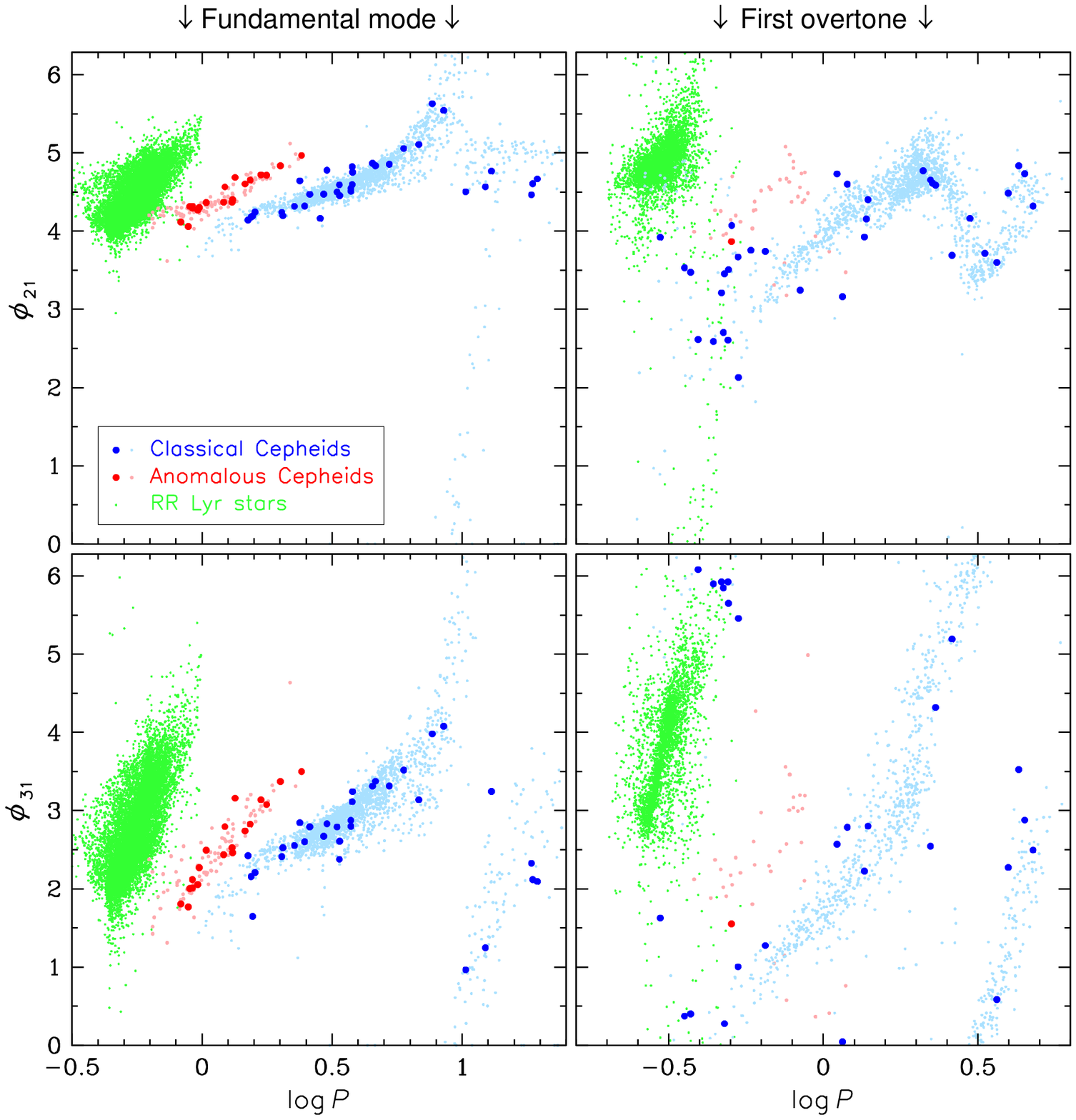}
\FigCap{Fourier coefficients $\phi_{21}$ and $\phi_{31}$ as a function of
  periods for single-mode classical Cepheids (blue points), anomalous
  Cepheids (red points), and RR~Lyr stars (green points) in the Galactic
  bulge (larger and darker points) and Large Magellanic Cloud (smaller and
  lighter points). {\it Left} and {\it right panels} show diagrams for
  fundamental-mode and first-overtone pulsators, respectively.}
\end{figure}

The Fourier coefficients were crucial for the first unambiguous
identification of anomalous Cepheids in the Galactic bulge. The complete
samples of classical pulsators in the Magellanic Clouds published by the
OGLE project (Soszyñski \etal 2015, 2017) allowed us to show that different
types of Cepheids are well separated in the period--$\phi_{21}$ and
period--$\phi_{31}$ diagrams. Fig.~1 shows these diagrams for the bulge
variables overplotted on their counterparts from the Large Magellanic Cloud
(LMC). The distinction between anomalous Cepheids and classical Cepheids
(and also RR~Lyr stars) is very prominent for the fundamental-mode
pulsators (shown in the left panels of Fig.~1), while for the
first-overtone pulsators (right panels of Fig.~1) the distinction is not so
clear, so our classification is less certain for these objects. We firmly
detected 20 anomalous Cepheids (nineteen fundamental-mode and one
first-overtone) in the Galactic bulge, eight of which were previously
classified as classical Cepheids (Soszyñski \etal 2011) and six as RR~Lyr
stars (Soszyñski \etal 2014). Previous designations of the reclassified
variables are given in the remarks of the collection.

Type~II Cepheids have traditionally been divided into BL~Her, W~Vir, and
RV~Tau stars based upon their pulsation periods. We used the same
discrimination periods as in the OGLE-III Catalog of Variable Stars
(Soszyñski \etal 2011) -- 5~d -- to distinguish between BL~Her and W~Vir
stars\footnote{Note that in the Magellanic Clouds (Soszyñski \etal 2008) we
  adopted a period of 4~d as a borderline between the BL~Her and W~Vir
  classes.} and 20~d to separate W~Vir and RV~Tau variables.

\begin{figure}[t]
\centerline{\includegraphics[width=10.8cm, bb=95 200 490 740]{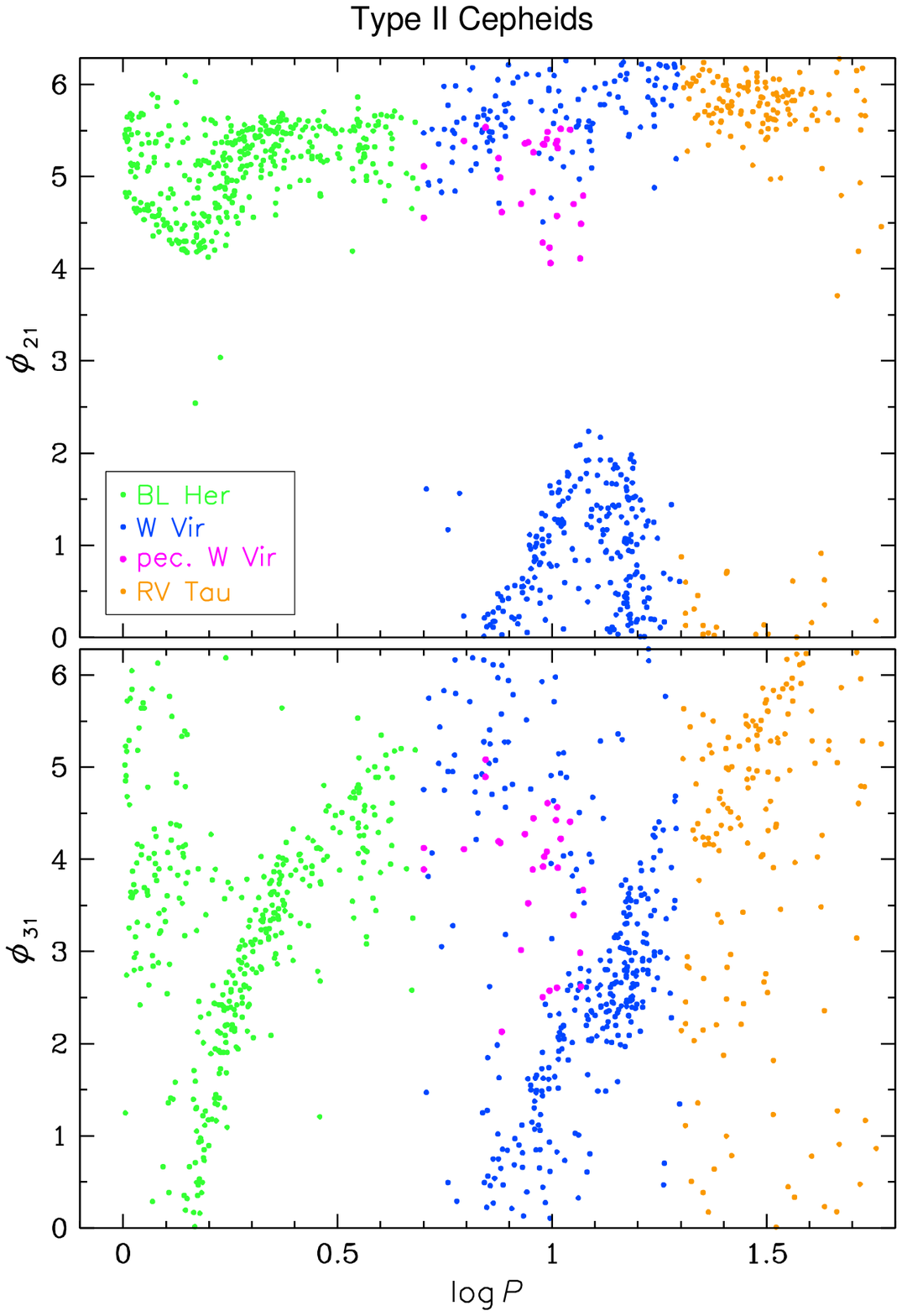}}
\FigCap{Fourier coefficients $\phi_{21}$ and $\phi_{31}$ as a function
of periods for type~II Cepheids in the Galactic bulge. Green points
indicate BL~Her stars, blue points -- regular W~Vir stars, magenta
points -- peculiar W~Vir stars, and orange points -- RV~Tau stars.}
\end{figure}

We also distinguished 30 candidates for peculiar W~Vir stars. This subclass
of type~II Cepheids was defined by Soszyñski \etal (2008) based on the
analysis of stars in the LMC. Peculiar W~Vir stars are on average brighter
and bluer than regular W~Vir stars, and have different light curve
morphology, with the rising branch being steeper than the declining
one. This latter feature is reflected in the period -- Fourier coefficients
diagrams (Fig.~2), where the peculiar W~Vir stars occupy a limited area,
although with some overlap with the regular W~Vir stars, so our
classification should be treated with caution. In the Galactic bulge, we
only found one peculiar W~Vir star exhibiting additional eclipsing
modulation (see Section 5.2), while in the Magellanic Clouds at least 30\%
of these pulsators show signs of binarity: eclipsing or ellipsoidal
variations.

\MakeTable{
l@{\hspace{8pt}} c@{\hspace{12pt}} | l@{\hspace{8pt}} c@{\hspace{6pt}}}{12.5cm}
{Reclassified stars from the OGLE-III Catalog of Variable Stars in the Galactic bulge}
{\hline \noalign{\vskip3pt}
\multicolumn{1}{c}{Identifier} & New            & \multicolumn{1}{c}{Identifier} & New            \\
                       & classification &                        & classification \\
\noalign{\vskip3pt}
\hline
\noalign{\vskip3pt}
OGLE-BLG-CEP-006   & Anom. Cepheid   & OGLE-BLG-T2CEP-245   & Other          \\
OGLE-BLG-CEP-008   & Anom. Cepheid   & OGLE-BLG-T2CEP-264   & Other          \\
OGLE-BLG-CEP-010   & Anom. Cepheid   & OGLE-BLG-T2CEP-295   & Other          \\
OGLE-BLG-CEP-011   & Type~II Cepheid & OGLE-BLG-T2CEP-323   & Other          \\
OGLE-BLG-CEP-012   & Other           & OGLE-BLG-RRLYR-08018 & Other          \\
OGLE-BLG-CEP-013   & Other           & OGLE-BLG-RRLYR-09254 & Class. Cepheid \\
OGLE-BLG-CEP-014   & Anom. Cepheid   & OGLE-BLG-RRLYR-10224 & Eclipsing      \\
OGLE-BLG-CEP-017   & Other           & OGLE-BLG-RRLYR-18443 & Anom. Cepheid  \\
OGLE-BLG-CEP-022   & Anom. Cepheid   & OGLE-BLG-RRLYR-21455 & Anom. Cepheid  \\
OGLE-BLG-CEP-023   & Anom. Cepheid   & OGLE-BLG-RRLYR-25667 & Anom. Cepheid  \\
OGLE-BLG-CEP-025   & Anom. Cepheid   & OGLE-BLG-RRLYR-27874 & Class. Cepheid \\
OGLE-BLG-CEP-026   & Artifact        & OGLE-BLG-RRLYR-28668 & Anom. Cepheid  \\
OGLE-BLG-CEP-028   & Anom. Cepheid   & OGLE-BLG-RRLYR-31394 & Spotted        \\
OGLE-BLG-T2CEP-042 & Other           & OGLE-BLG-RRLYR-33138 & Anom. Cepheid  \\
OGLE-BLG-T2CEP-070 & Other           & OGLE-BLG-RRLYR-36100 & Anom. Cepheid  \\
OGLE-BLG-T2CEP-099 & Other           & OGLE-BLG-ECL-297887  & Type~II Cepheid \\
\noalign{\vskip3pt}
\hline}

Our analysis revealed 68 classical Cepheids, 574 type~II Cepheids and 828
RR~Lyr stars that were not included in the previous versions of the
OCVS. Some objects have been reclassified which is summarized in
Table~1. In addition to the aforementioned pulsators that were reclassified
as anomalous Cepheids, we removed from the collection several objects which
likely belong to other (usually unknown) types of variable stars. We
confirm that all the five stars in the flared disk of the Milky Way studied
by Feast \etal (2014) are indeed the fundamental-mode classical Cepheids.

\Section{OGLE Collection of Cepheids in the Galactic Bulge}
The newly detected classical pulsators have been added to the previously
published OGLE catalogs. In total, the OCVS now contains 87 classical
Cepheids, 924 type~II Cepheids, 20 anomalous Cepheids, and 39\,074 RR~Lyr
stars in the Galactic bulge. The entire collection can be downloaded {\it
  via} the WWW interface or from the FTP site:
\begin{center}
{\it http://ogle.astrouw.edu.pl}\\
{\it ftp://ftp.astrouw.edu.pl/ogle/ogle4/OCVS/blg/cep/}\\
{\it ftp://ftp.astrouw.edu.pl/ogle/ogle4/OCVS/blg/t2cep/}\\
{\it ftp://ftp.astrouw.edu.pl/ogle/ogle4/OCVS/gal/acep/}\\
{\it ftp://ftp.astrouw.edu.pl/ogle/ogle4/OCVS/blg/rrlyr/}
\end{center}

Each object in our collection received a unique identifier which, with some
exceptions, follows the scheme proposed in the previous parts of the
OCVS. For example, classical Cepheids are designated as OGLE-BLG-CEP-NNN,
where NNN is a three-digit number. In the OGLE-III catalog (Soszyñski \etal
2011), we used two-digit numbers in the identifiers of classical Cepheids,
but in this work we reached number OGLE-BLG-CEP-100, which forced us to
slightly change the naming scheme. Objects from OGLE-BLG-CEP-001 to
OGLE-BLG-CEP-032 were included in the OGLE-III Catalog of Variable Stars
(Soszyñski \etal 2011), while objects from OGLE-BLG-CEP-033 to
OGLE-BLG-CEP-100 are the newly discovered classical Cepheids. These new
pulsators are ordered by increasing right ascension. The names of anomalous
Cepheids -- OGLE-GAL-ACEP-NNN -- follow the scheme proposed by Soszyñski
\etal (2017), who discovered seven Galactic anomalous Cepheids in front of
the Magellanic Clouds.

All tables on the FTP site are in ASCII format and contain basic
information about the stars: their pulsation modes, J2000 equatorial
coordinates, intensity mean magnitudes in the {\it I}- and {\it V}-bands,
periods in days with their uncertainties (derived with the {\sc Tatry} code
of Schwarzenberg-Czerny 1996), epochs of maximum light, peak-to-peak
amplitudes in the {\it I}-band, and Fourier coefficients $R_{21}$,
$\phi_{21}$, $R_{31}$, $\phi_{31}$ derived for the {\it I}-band light
curves. For the already known variables we provide their identifiers from
the International Variable Star Index (Watson \etal 2006).

\Section{Discussion}
\Subsection{Classical Cepheids}
Fig.~3 shows the sky distribution of the OGLE classical Cepheids in the
Galactic coordinates. Despite the fact that classical Cepheids are young
($<300$~Myr), relatively massive (3.5--20~\MS) stars, a large fraction
of our objects are located far from the Galactic plane, up to the Galactic
latitudes $|b|=5\arcd$. Feast \etal (2014) measured distances to five OGLE
fundamental-mode Cepheids and showed that they are located in the flared
outer disk of the Milky Way, from 0.9~kpc to 2.1~kpc from the Galactic
plane. Our collection probably contains more classical Cepheids in the
flared disk, not necessarily only the fundamental-mode pulsators.
\begin{figure}[htb]
\begin{center}
\includegraphics[width=11.6cm]{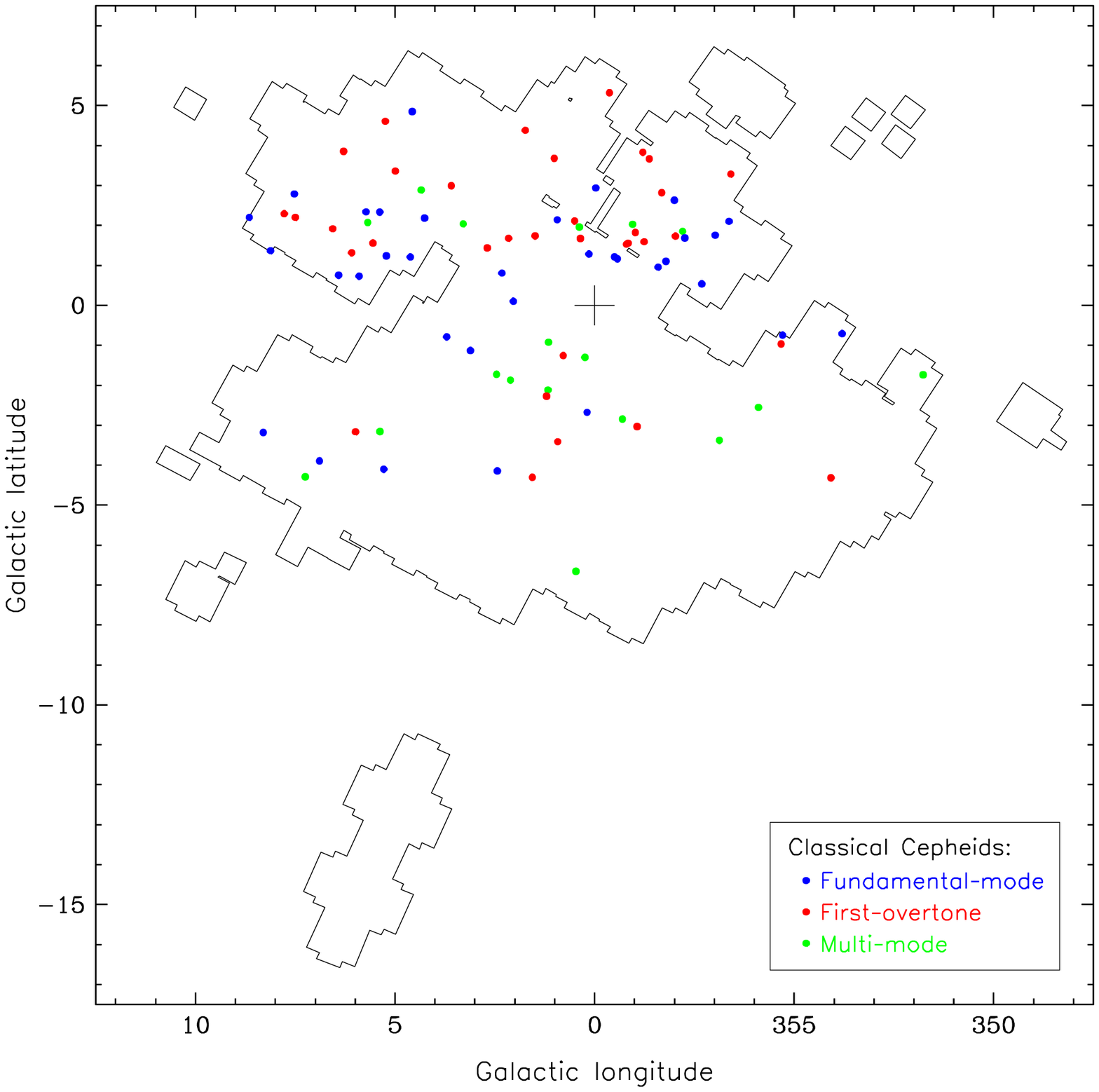}
\end{center}
\FigCap{Positions in the sky of the OGLE classical Cepheids (in
Galactic coordinates). Blue, red and green points mark
fundamental-mode, first-overtone, and multi-mode pulsators,
respectively. Black contours show borders of the OGLE fields.}
\end{figure}

The presence of classical Cepheids inside the Galactic bulge is currently a
matter of debate. The VVV survey reported the discovery of 35 classical
Cepheids in the highly-obscured regions of the Galactic bulge
(D{\'e}k{\'a}ny \etal 2015) and suggested that the young stellar population
form an inner thin disk surrounding the Galactic center. Matsunaga \etal
(2016) used the IRSF/SIRIUS near-infrared observations to detect their own
set of Cepheids in this area (partly overlapping with the VVV Cepheids),
but they concluded that almost all of these variables are located behind
the bulge and the Galactic center lacks classical Cepheids. The only
exception from this rule are four classical Cepheids discovered by
Matsunaga \etal (2011, 2015) in the very center of the Milky Way, in the so
called Nuclear Bulge or Central Molecular Zone.

We cannot participate in this interesting discussion, because the Nuclear
Bulge is inaccessible to optical observations owing to enormous
interstellar extinction. However, the OGLE Collection of Variable Stars
also contains several classical Cepheids relatively close ($|b|<1\arcd$) to
the plane of the Milky Way. These objects are probably located in the
Galactic disk in front of the Galactic bulge. For most of these stars, the
{\it V}-band data are not available, usually because of the high reddening
which shifts the {\it V}-band magnitudes below the OGLE detection
threshold.

Our collection includes 16 double-mode and three triple-mode classical
Cephe\-ids. Positions of these stars in the Petersen diagram
(shorter-to-longer period ratio \vs the longer period) overplotted on the
LMC and SMC beat Cepheids are shown in Fig.~4. Soszyñski \etal (2011)
showed that period ratios of double-mode classical Cepheids in the Galactic
bulge are smaller than in the Magellanic Clouds, which is probably caused
by metallicity differences between these stellar environments. Our extended
sample confirms that indeed double-mode Cepheids in the bulge, both F/1O
and 1O/2O pulsators, have smaller period ratios than their counterparts in
the Magellanic Clouds, but this rule is valid only for the short-period
variables. The longest period beat Cepheids (two F/1O and two 1O/2O
pulsators) have period ratios virtually the same as variables in the
Magellanic Clouds.

\begin{figure}[htb]
\centerline{\includegraphics[width=11.5cm]{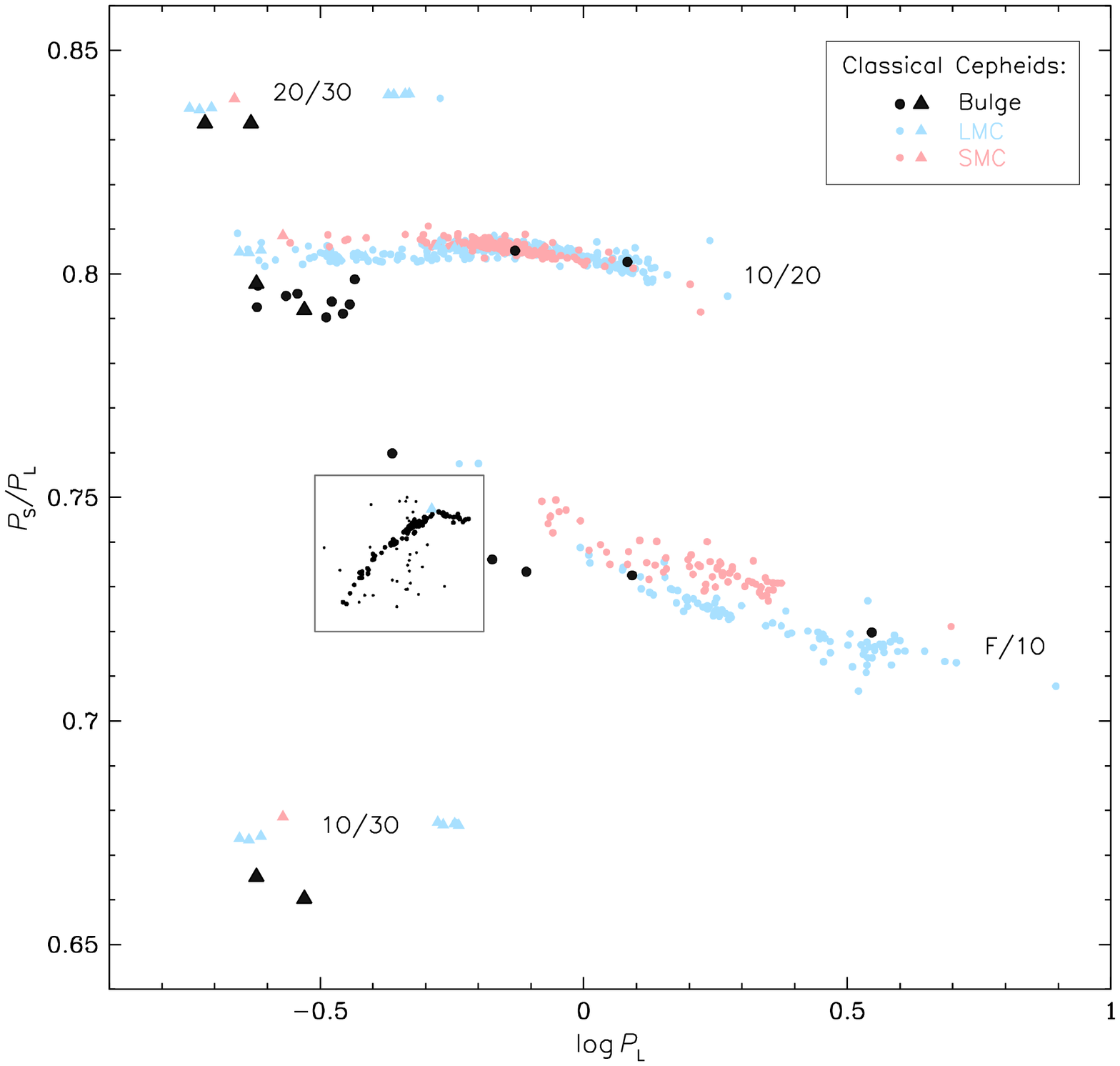}}
\FigCap{Petersen diagram for double-mode (circles) and triple-mode
(triangles) classical pulsators detected by the OGLE project toward
the Galactic bulge (black symbols), LMC (light blue symbols), and SMC
(light red symbols). Gray rectangle separates multi-mode RR~Lyr stars
-- this region is zoomed in Fig.~11.}
\end{figure}
Triple-mode Cepheids (marked with triangles in Fig.~4) have generally
short periods and also tend to have smaller period ratios compared to
the LMC and SMC variables.

\Subsection{Type~II Cepheids}
In this paper, we present the most numerous sample of type~II Cepheids
detected in one stellar environment. Two-dimensional spatial
distribution of 924 type~II Cepheids from our collection is presented
in Fig.~5. These stars generally show a strong concentration toward
the Galactic center, but they avoid regions around the Galactic
plane. This is of course caused by the large amount of interstellar
matter toward these regions which obscures stars in the optical
regime. Such a distribution suggests that the vast majority of our
type~II Cepheids are members of the Galactic bulge.
\begin{figure}[htb]
\centerline{\includegraphics[width=12cm]{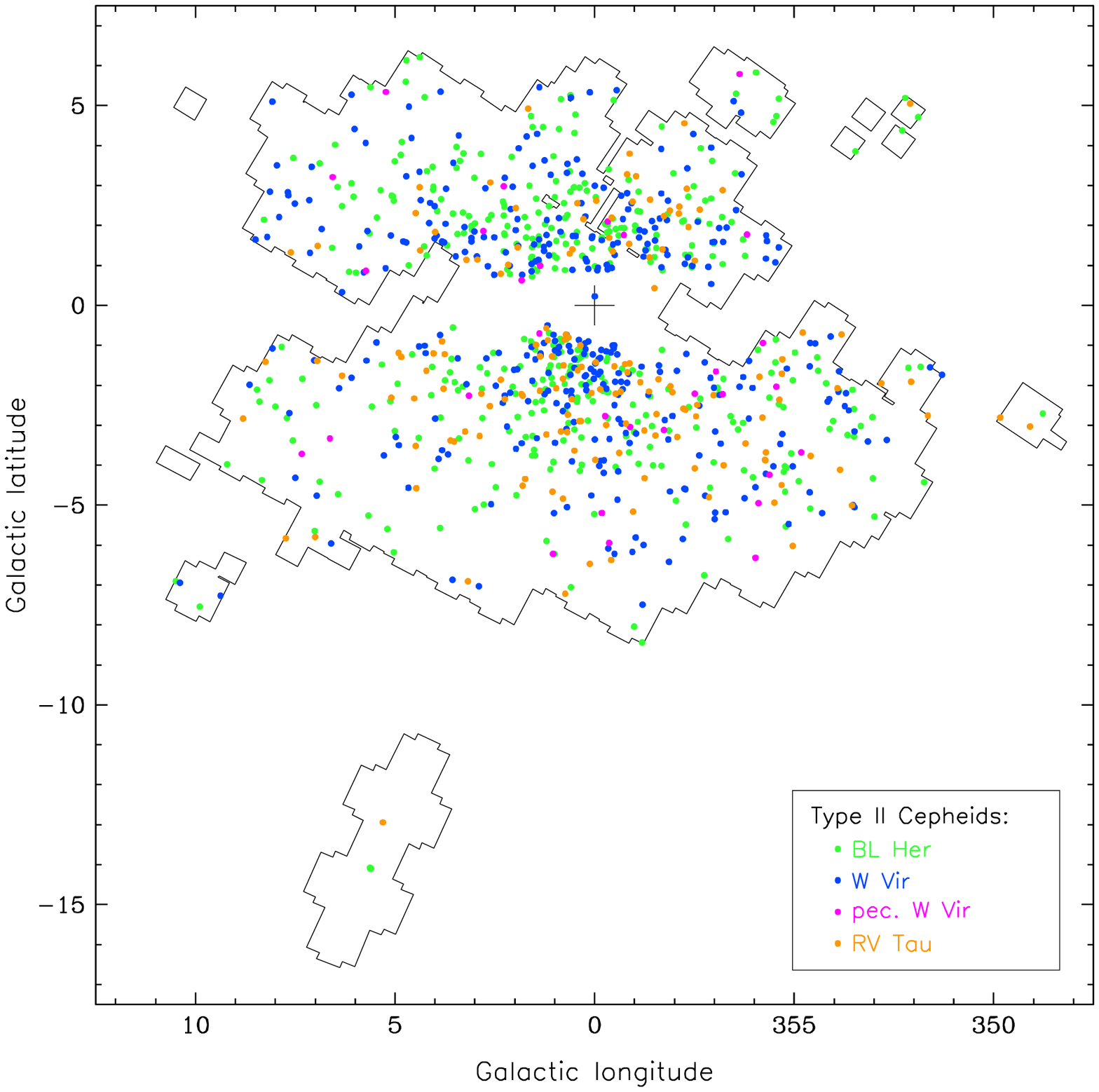}}
\FigCap{Positions in the sky of the OGLE type~II Cepheids (in Galactic
coordinates). Green, blue, magenta, and orange points mark BL~Her,
W~Vir, peculiar W~Vir, and RV~Tau stars, respectively. Black contours
show borders of the OGLE fields.}
\end{figure}

\begin{figure}[htb]
\centerline{\includegraphics[width=12.5cm, bb=25 60 545 745]{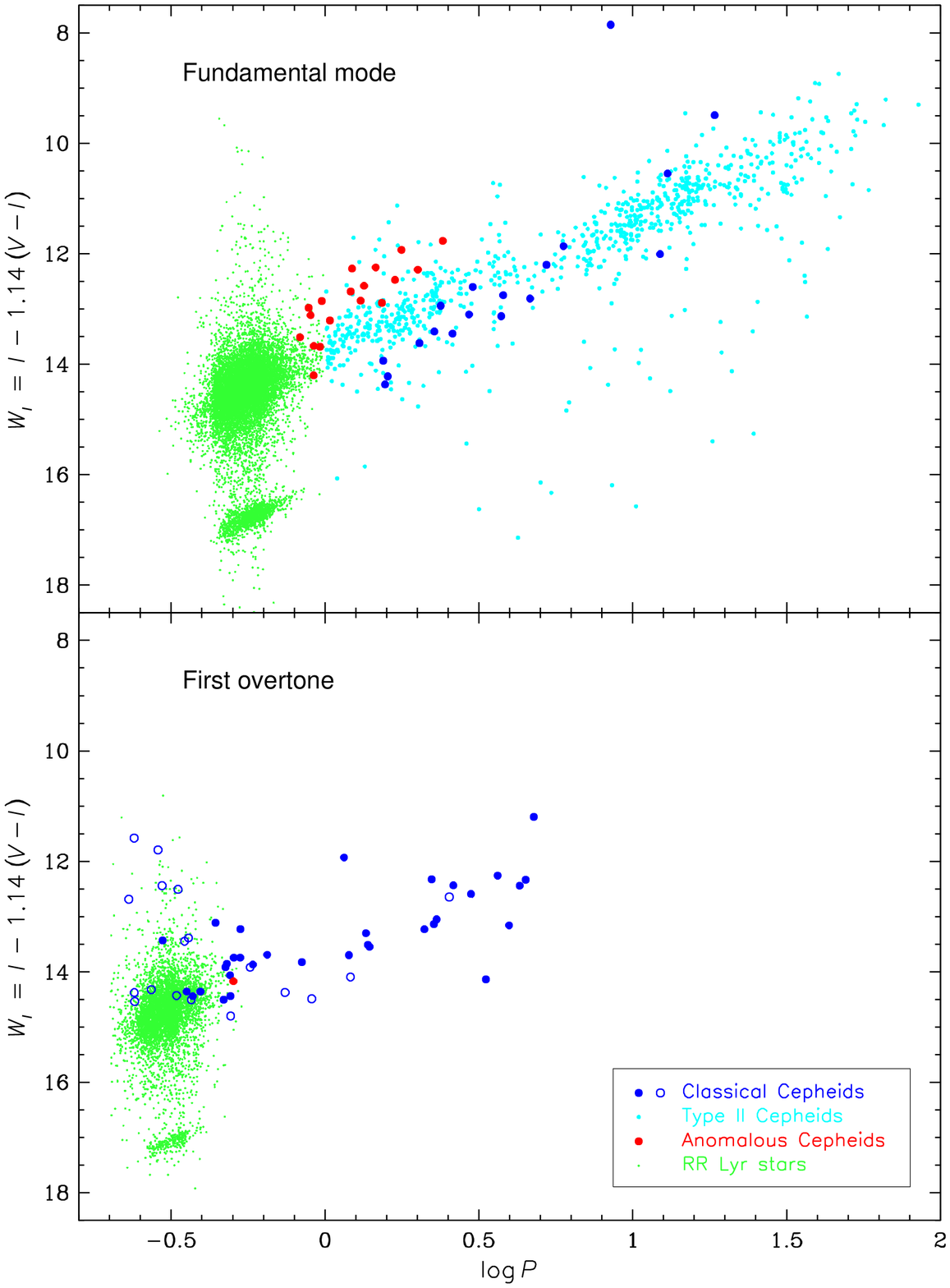}}
\FigCap{Period \vs Wesenheit index diagram for classical Cepheids (blue
points), type~II Cepheids (cyan points), anomalous Cepheids (red
points), and RR~Lyr stars (green points). {\it Upper panel} shows
fundamental-mode variables, while {\it lower panel} presents first
overtone pulsators. Open blue circles mark the first-overtone periods
of the multi-mode Cepheids.}
\end{figure}

The same conclusion can be drawn from the analysis of the PL diagram
(Fig.~6), where as the ``luminosity'' we used the reddening-independent
Wesenheit index in the Galactic bulge defined by Pietrukowicz \etal (2015)
as $W_I=I-1.14(V-I)$ (Fig.~6 includes only those variables, for which both,
{\it I}- and {\it V}-band, mean magnitudes are available). The majority of
type~II Cepheids populate a PL relation that is an extension of the
relation for RRab stars in the bulge, which indicates that both classes are
on average at the same distance from us. In turn, most of the classical
Cepheids are fainter than type~II Cepheids with the same periods, which
suggests that the most classical Cepheids in our sample are located far
behind the Milky Way center, also in the flared disk (Feast \etal 2014). As
already mentioned, a number of classical Cepheids in our sample lie in
front of the Galactic bulge, but most of them do not have {\it V}-band
measurements and they are not included in the period \vs Wesenheit index
diagram.

Our collection covers the full range of periods observed in type~II
Cepheids: from 1~d to 66~d\footnote{In the OCVS we provide ``single''
periods, \ie intervals between successive minima, even if the
period-doubling effect is present}. BL~Her variables with the shortest
periods are adjacent to the longest-period RR~Lyr stars and the
boundary period between these groups is not strictly defined. We
traditionally used a 1.0~d period as a borderline between RR~Lyr and
BL~Her stars.

\begin{figure}[htb]
\includegraphics[width=12.5cm, bb=20 600 565 760]{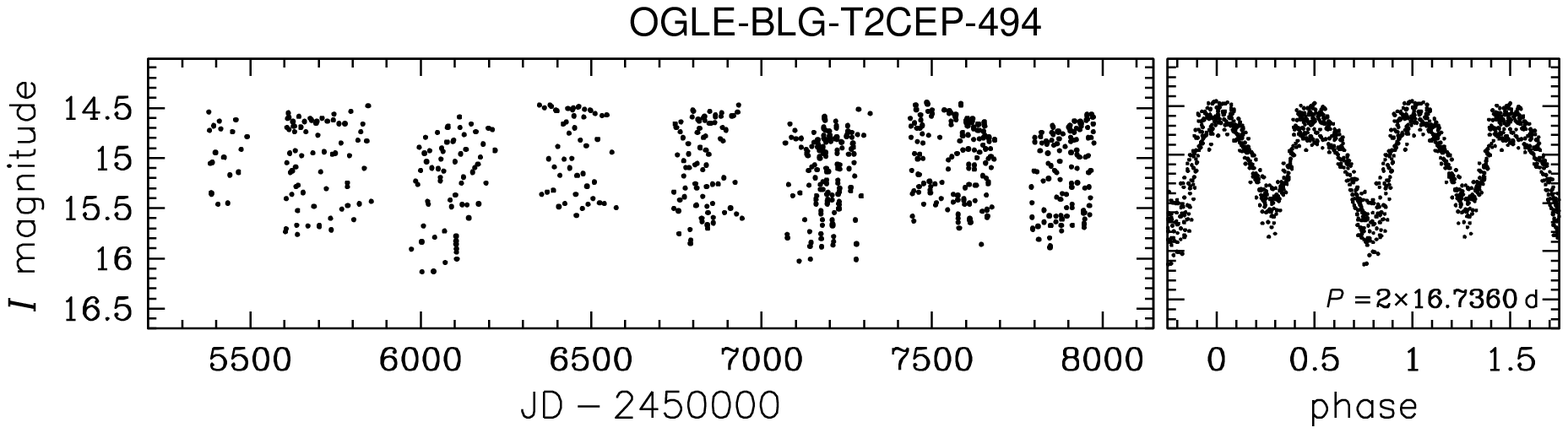}
\FigCap{{\it I}-band light curve of OGLE-BLG-T2CEP-494 -- one of the
shortest-period type~II Cepheid with a prominent RV-Tau-like behavior:
periodic alternations of deep and shallow minima and the long-term
variations of the mean brightness (RVb type). {\it Left panel}
presents unfolded light curve collected in the years 2010--2017. {\it
Right panel} shows the same light curve folded with the doubled
pulsation period.}
\end{figure}

In our collection, type~II Cepheids with pulsation periods longer than 20~d
are classified as RV~Tau stars. The characteristic feature of RV~Tau
variables -- alternating deep and shallow minima (period doubling) -- seems
not to be a distinctive classification characteristic. On the one hand, the
period doubling is a common phenomenon also among long-period W~Vir stars,
down to a period of about 16~d. Fig.~7 shows the light curve of
OGLE-BLG-T2CEP-494 ($P=16.736$~d) which exhibits alternations of minima and
maxima. On the other hand, our collection contains some long-period type~II
Cepheids ($P>20$~d) with no clear alternations of minima, but in our
catalog they are formally categorized as RV~Tau stars. Such stars may also
be classified as yellow semiregular variables (SRd stars).

At least 15 RV~Tau variables in our collection belong to the RVb class,
which means that these stars experience long-term variations of the mean
brightness, with periods of 470--2800~d and amplitudes from 0.2~mag to
2.5~mag. Also two W~Vir stars (with pulsation period shorter than 20~d)
exhibit such modulation (Fig.~7). The long-term changes are commonly
interpreted as being caused by periodic obscuration of a binary system by a
circumbinary dust disk (\eg Evans 1985, Pollard \etal 1996, Van
Winckel \etal 1999). The long-term homogeneous OGLE light curves in two
standard filters may provide important constraints on theoretical models of
the RVb phenomenon. In some objects, the depths of the RVb modulation
significantly change from cycle to cycle.

The OGLE catalog of type~II Cepheids in the Magellanic Clouds (Soszyñski
\etal 2008, 2010) contains an exceptionally large fraction of pulsators
that are members of binary systems. Over a dozen type~II Cepheids show
additional eclipsing or ellipsoidal modulation. Most of these objects were
classified as peculiar W~Vir stars, which suggests that the binarity might
be related to the origin of this class of pulsating stars (see the
discussion in Pilecki \etal 2017).

\begin{figure}[htb]
\includegraphics[width=12.5cm, bb=25 225 585 775]{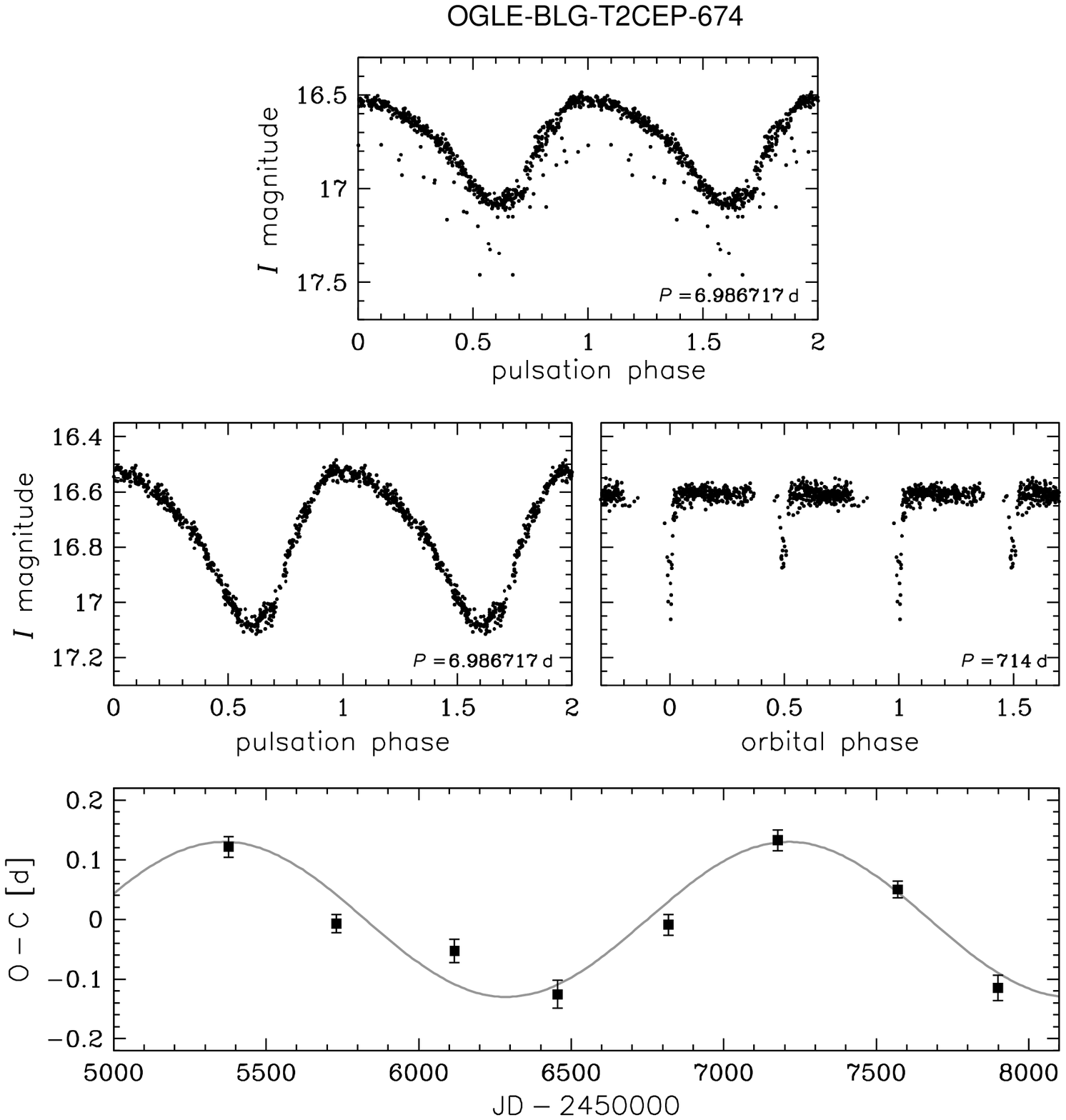}
\FigCap{OGLE-BLG-T2CEP-674 -- a type~II Cepheid with additional
eclipsing variability. {\it Upper panel} shows the original light
curve folded with the pulsation period, {\it middle left panel}
presents the pulsation light curve after removing the eclipses, {\it
middle right panel} presents the eclipsing light curve prewhitened
with the pulsations, {\it lower panel} shows the $O-C$ diagram
obtained in the years 2010--2017.}
\end{figure}
It appears that the fraction of eclipsing Cepheids in the Galactic bulge is
much lower than in the Magellanic Clouds. In our collection we found only
one type~II Cepheid that exhibits additional eclipsing modulation:
OGLE-BLG-T2CEP-674. Its light curve is shown in the upper panel of Fig.~8,
while middle panels show disentangled pulsation and eclipsing
variabilities. The orbital period of this system is close to 2~years
(714~d), which interferes with annual gaps in the OGLE photometry, but
luckily both, primary and secondary, eclipses are well visible in the light
curve.

A closer look at the photometric data of OGLE-BLG-T2CEP-674 reveals
yet another interesting feature. Based on the pulsation light curve we
constructed an observed-minus-calculated ($O-C$) diagram which is
presented in the lower panel of Fig.~8. This diagram shows a long-term
sinusoidal-like variations which may be caused by the light-travel
time effect in a binary system. Surprisingly, a possible period of the
sinusoidal variations visible in the $O-C$ diagram (about 1900~d) is
much longer than the orbital period measured from the eclipsing
modulation (714~d), which suggests that OGLE-BLG-T2CEP-674 is a member
of at least a triple system with the third body on a $\sim$1900~d
orbit. However, we must stress here that this period is comparable to
the total time span of the OGLE-IV photometry (the star was not
observed during the previous phases of the OGLE survey) so we cannot
be sure that the variations in the $O-C$ diagram are indeed periodic
ones.

\Subsection{Anomalous Cepheids}
In this paper, we present the first {\it bona fide} anomalous Cepheids
detected in the Galactic bulge. Eight of these stars were published by
Soszyñski \etal (2011) as classical Cepheids and six other objects were
previously classified as RR~Lyr stars (Soszyñski \etal 2014), because at
that time there was no reliable method to distinguish anomalous Cepheids
from other types of classical pulsators based solely on their light curve
shape. This situation changed when Soszyñski \etal (2015, 2017) selected
the richest known population of anomalous Cepheids in the Magellanic Clouds
and showed that the shape of their light curves (quantitatively assessed by
the Fourier coefficients $\phi_{21}$ and $\phi_{31}$) is an efficient
diagnostic to separate these groups. Positions in the sky of the 20
anomalous Cepheids in the Galactic bulge are shown in Fig.~9. Despite a
small number of objects, we may notice a lack of anomalous Cepheids close
to the Galactic plane due to high interstellar extinction in these regions.
\begin{figure}[b]
\centerline{\includegraphics[width=10.7cm]{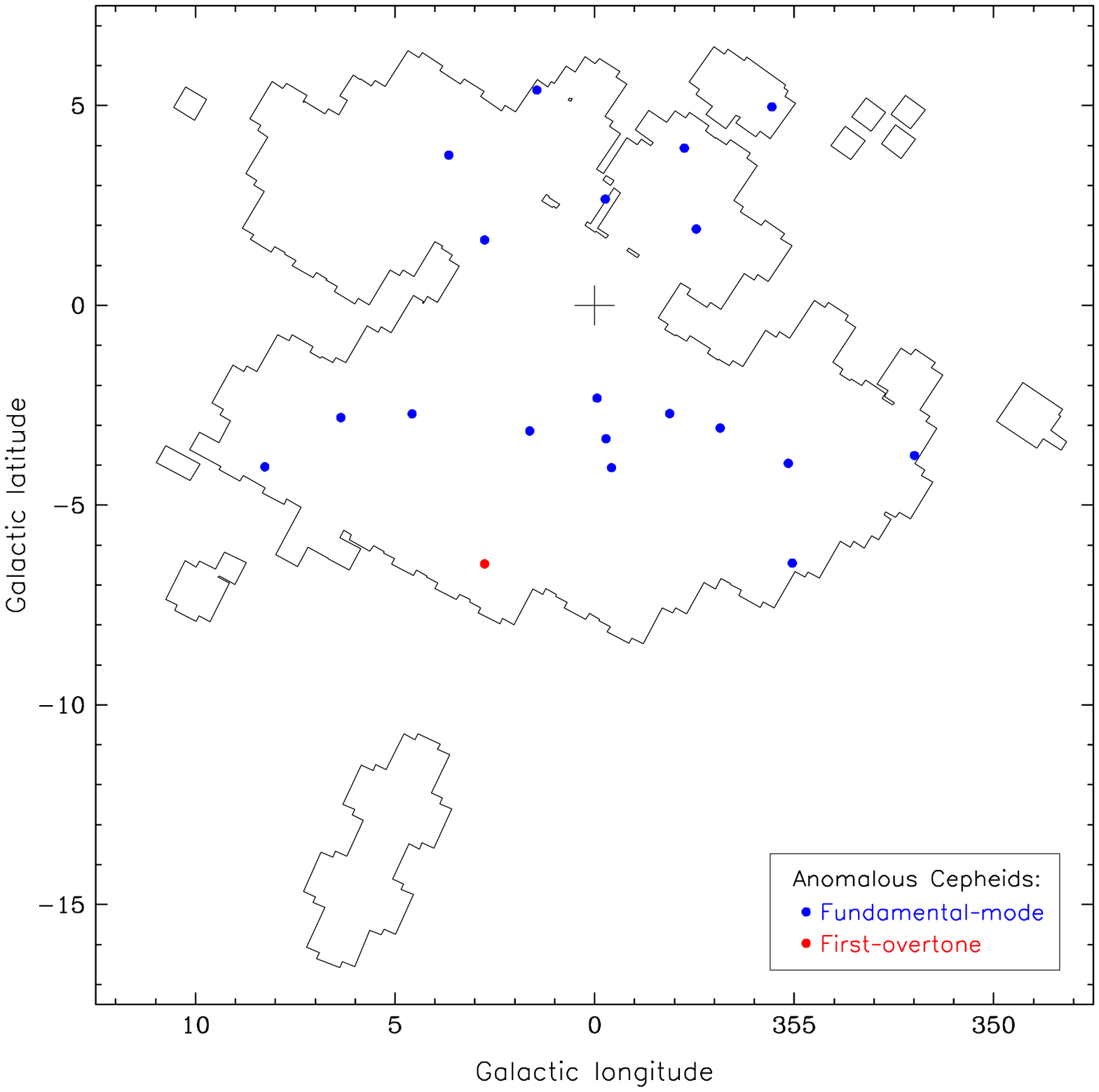}}
\FigCap{Positions in the sky of the OGLE anomalous Cepheids (in
Galactic coordinates). Blue and red symbols mark fundamental-mode and
first-overtone pulsators, respectively. Black contours show borders of
the OGLE fields.}
\end{figure}

Fundamental-mode anomalous Cepheids are well separated from classical
Ce\-pheids in the PL diagram (Fig.~6), but, paradoxically, anomalous
Cepheids have brighter apparent magnitudes than their classical
siblings. An analysis of pulsating stars in the LMC (Soszyñski \etal 2015)
shows that anomalous Cepheids are intrinsically fainter by about 0.7~mag
than classical Cepheids. On the other hand, anomalous Cepheids are on
average brighter than type~II Cepheids, just like in the Magellanic Clouds,
so we can safely assume that our sample of anomalous Cepheids belongs to
the Galactic bulge.

\Subsection{RR~Lyr stars}
The newly detected 828 RR~Lyr stars represent about 2\% of all variables of
this type discovered by OGLE in the central regions of the Galaxy
(Soszyñski \etal 2014). These new detections were previously overlooked for
various reasons. Most of the new RR~Lyr variables (72\%) are first-overtone
pulsators (RRc stars) which usually show nearly sinusoidal light curves
that can be confused with other classes of variable stars, for example
close binary systems. Some variables were missed because of a small number
of data points, or very low amplitudes, or pulsation periods close to 1 or
2/3 of the sidereal day which affected the period determinations.

\begin{figure}[b]
\centerline{\includegraphics[width=12cm, bb=25 365 575 765]{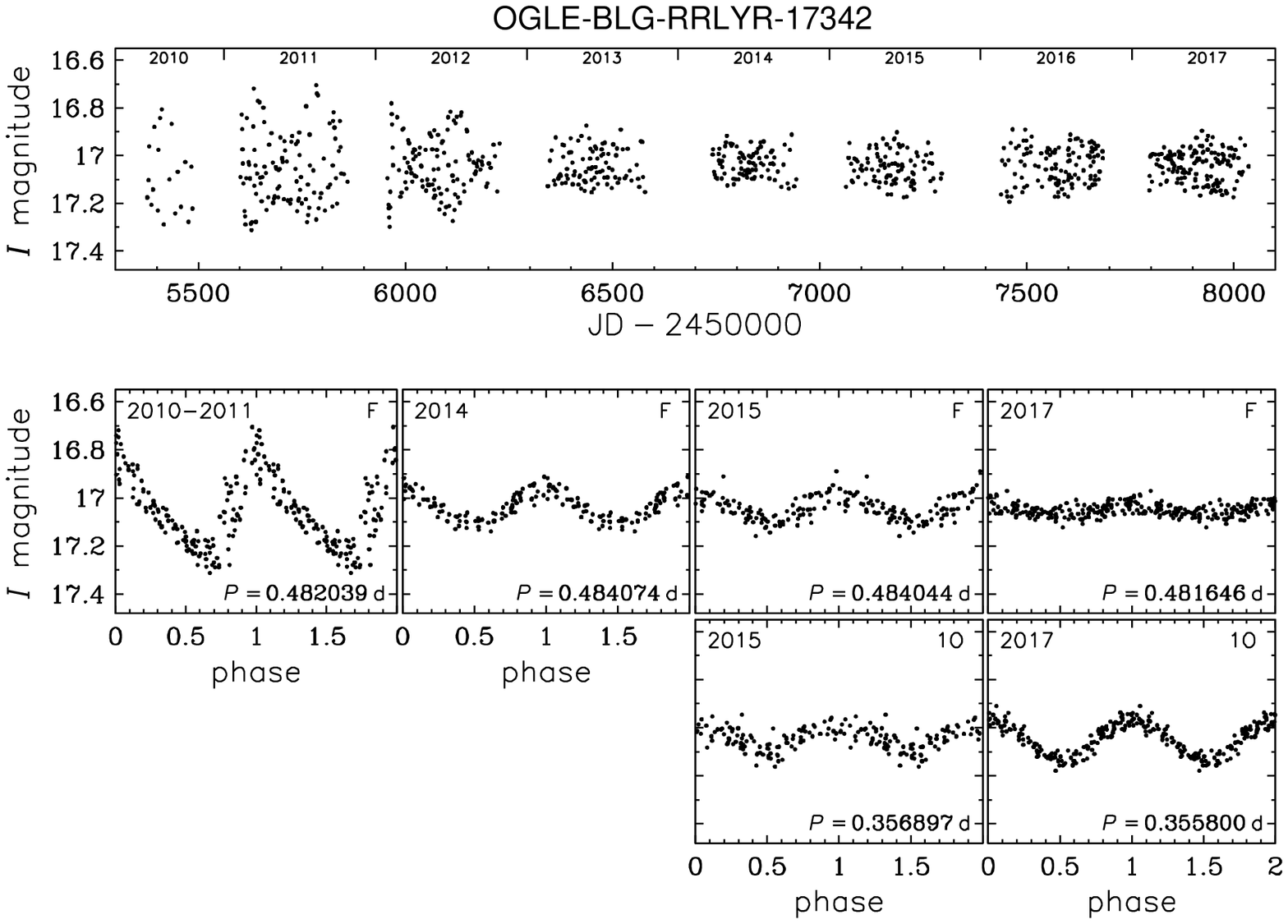}}
\FigCap{{\it I}-band light curve of OGLE-BLG-RRLYR-17342 -- an RR~Lyr
star that switched from a single-mode (RRab) to double-mode (anomalous
RRd) pulsation. {\it Upper panel} shows unfolded OGLE-IV light curve
collected in the years 2010--2017. {\it Lower panels} show folded light
curves obtained in the selected seasons: 2010--2011, 2014, 2015, and
2017. From 2015, when the first-overtone mode turned on, we present
disentangled light curves of both modes. Note changes of the pulsation
periods in the different seasons.}
\end{figure}
The time-series photometry of RR~Lyr stars published by Soszyñski
\etal (2014) have been supplemented with new observations collected by
the OGLE-IV survey up to August 2017. At present, the time-span of the
OGLE-IV light curves exceeds 7 years and for some stars it can be
increased to even 20 years by joining the OGLE-II and OGLE-III data
points. These light curves can be used for follow-up studies of all
stationary and non-stationary phenomena in pulsating stars: non-radial
modes, Blazhko effect, period changes, mode switching, light-time
effect in binary system hosting pulsating stars, etc.

Fig.~10 shows an example of such non-stationary behaviors: the light
curve of OGLE-BLG-RRLYR-17342 -- a Blazhko RR~Lyr star that switched
from a single-mode RRab star to a double-mode RRd star. Lower panels
of Fig.~10 display folded light curves of OGLE-BLG-RRLYR-17342
obtained by OGLE in selected seasons. The amplitude of the
fundamental-mode pulsation has continuously decreased over the eight
years of the monitoring and at present it is slightly above to the
detection limit of the OGLE photometry. In 2015, the first-overtone
mode appeared in the light curve and the star became a double-mode
pulsator. Three other mode-switching RR~Lyr stars in the Galactic
bulge were reported by Soszyñski \etal (2014).
\begin{figure}[b]
\centerline{\includegraphics[width=10.7cm]{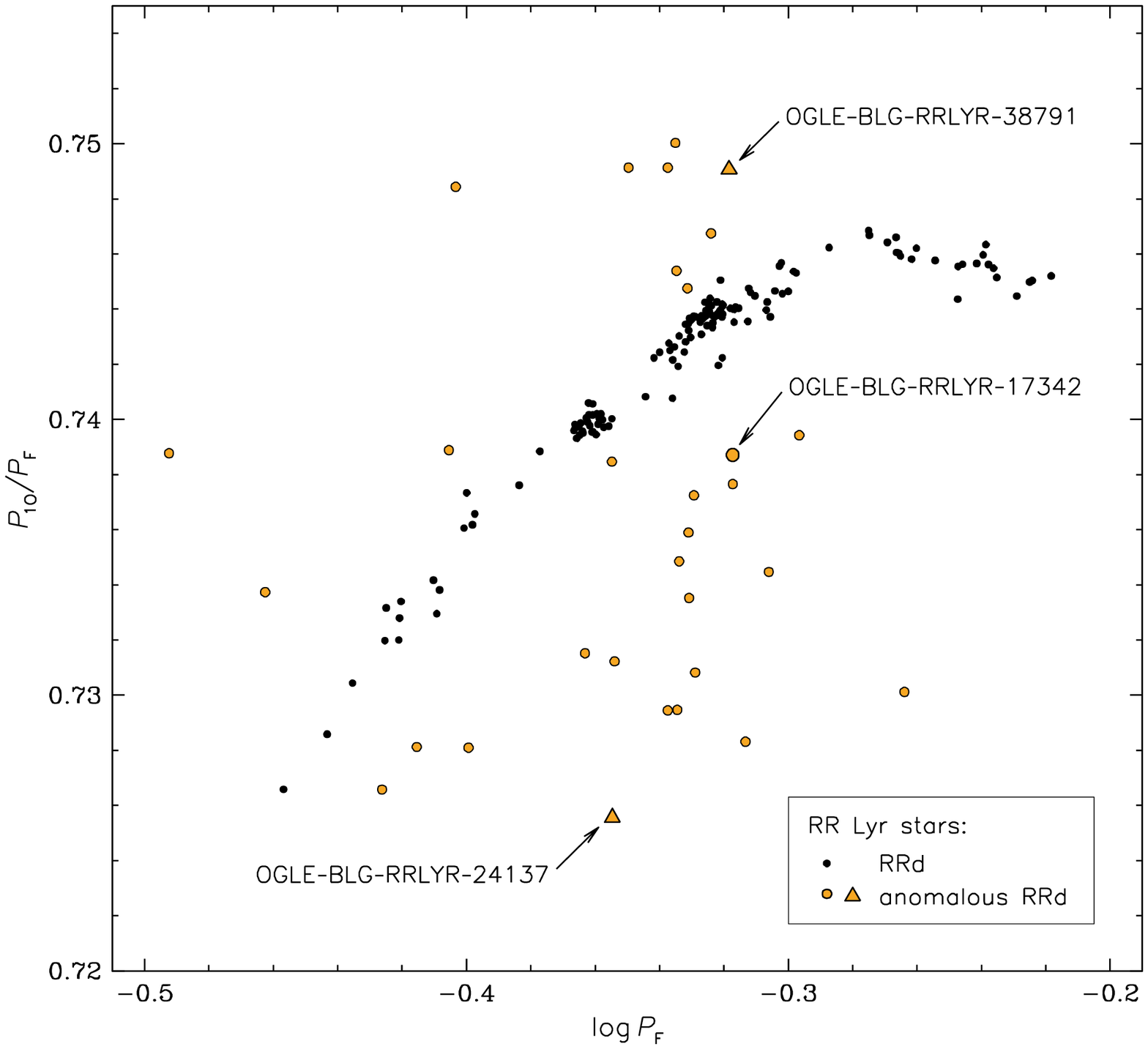}}
\vskip5pt
\FigCap{Petersen diagram for double-mode (circles) and triple-mode
(triangles) RR~Lyr stars in the Galactic bulge. Black points mark
``regular'' RRd stars, yellow symbols represent anomalous RRd stars
(Soszyñski \etal 2016b). Triple-mode stars -- OGLE-BLG-RRLYR-24137 and
OGLE-BLG-RRLYR-38791 -- and a mode-switching pulsator --
OGLE-BLG-RRLYR-17342 -- are indicated by arrows.}
\end{figure}

We reclassified 24 double-mode RR~Lyr stars from ordinary to anomalous
RRd stars. This latter class of pulsators was defined by
Soszyñski \etal (2016b) who noticed a distinct group of double-mode
RR~Lyr variables in the Magellanic Clouds. Anomalous RRd stars have
different period and amplitude ratios than typical RRd stars and most
of them show Blazhko modulation (Smolec \etal 2015a). In the present
investigation, the main classification criterion was the position of a
given pulsator in the Petersen diagram (Fig.~11). All multi-mode
RR~Lyr variables that are located outside the curved sequence in the
Petersen diagram (both, above and below, this sequence) were
classified as anomalous RRd stars. Thus, we have expanded the
definition of anomalous double-mode RR~Lyr variables introduced by
Soszyñski \etal (2016b) for the Magellanic Clouds members by adding
objects with the $P_{\textrm{1O}}/P_\textrm{F}$ period ratios higher
than observed for ``classical'' RRd stars. The majority of our
candidates for anomalous RRd stars in the Galactic bulge share the
features of their Magellanic Clouds counterparts: usually the
fundamental mode is the dominant one and most of these objects exhibit
the Blazhko effect. Currently, the OCVS contains 31 anomalous RRd
stars, both reclassified and newly discovered variables. This group
includes also two mode-switching RR~Lyr stars: OGLE-BLG-RRLYR-13442
and OGLE-BLG-RRLYR-17342.

\begin{figure}[htb]
  \centerline{\includegraphics[width=12.0cm]{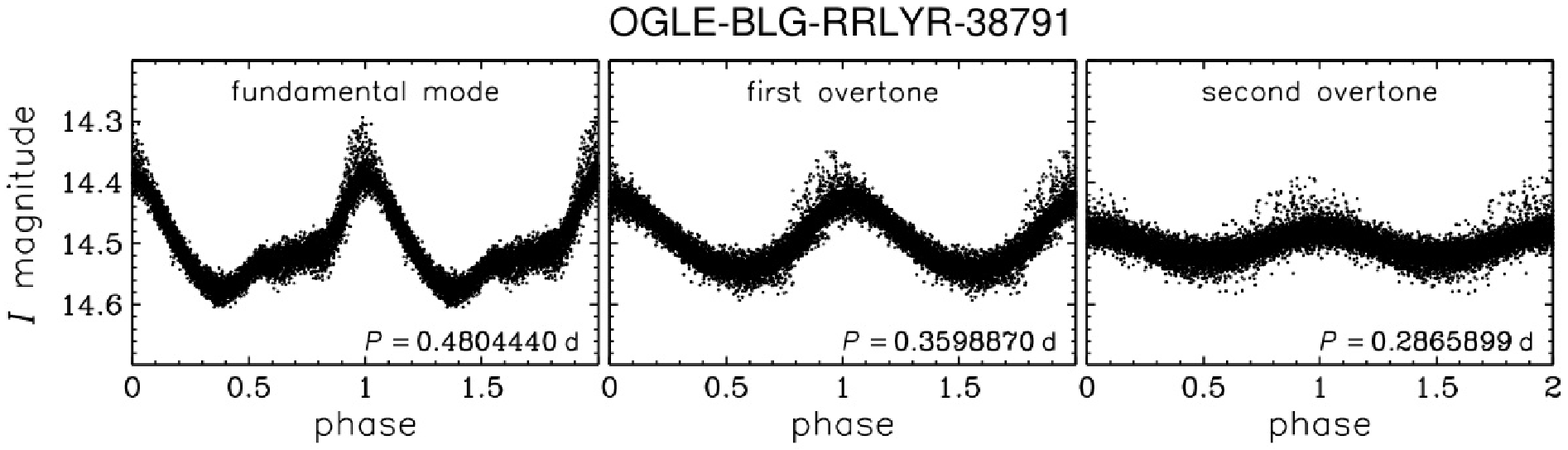}}
  \FigCap{{\it I}-band light curve of OGLE-BLG-RRLYR-38791 -- a triple-mode
    anomalous RRd star. {\it Each panel} shows a folded light curve of this
    star prewhitened with the two other pulsation modes.}
\end{figure}
One of the new detections deserves special attention.
OGLE-BLG-RRLYR-38791 is a triple-mode star pulsating likely in the
fundamental, first-overtone, and second-overtone modes. Its
disentangled light curves corresponding to the three pulsation modes
are displayed in Fig.~12. The classification of this object is
unclear. We decided to list it among anomalous RRd stars for a few
reasons. First, its fundamental-mode and first-overtone periods place
this object among anomalous RRd pulsators in the Petersen diagram
(Fig.~11). Second, the fundamental-mode has the largest amplitude. And
third, the shape of the fundamental-mode light curve resembles those
in some anomalous RRd stars.

Another triple-mode RR~Lyr star was found in the OCVS by Smolec \etal
(2015b). OGLE-BLG-RRLYR-24137 (in the present work also classified as an
anomalous RRd star) exhibits two radial fundamental and first-overtone
modes and the third periodicity that may correspond to the radial
third-overtone mode or a non-radial mode.

\Section{Conclusions}
We present the largest collection of classical, type~II, and anomalous
Cepheids in and toward the central regions of the Milky Way. We release the
long-term time-series photometry obtained by the OGLE project for all the
stars. The OGLE samples of classical pulsators provide us with a tool to
test evolutionary and stellar pulsation models, as well as to study the
structure and star formation history in the Galactic bulge region, which is
crucial for our understanding of the Milky Way evolution.

In the near future, the OCVS will be extended by classical pulsators found
within the OGLE Galaxy Variability Survey regularly observing an area of
about 2000 square degrees in the Galactic disk and outer regions of the
Galactic bulge. This sub-project of the OGLE survey has been carried out
since 2013 and currently it accumulated long-term multi-epoch photometric
data which can be used to perform an effective search for variable stars
with their reliable classification. This dataset promises to be a powerful
tool to improve our understanding of the Galactic structure.

\Acknow{We would like to thank Profs. M.~Kubiak and G.~Pietrzyñ\-ski,
  former members of the OGLE team, for their contribution to the collection
  of the OGLE photometric data over the past years. We are grateful for
  discussions and constructive comments to R.~Smolec.  We thank
  Z.~Ko³aczkowski and A.~Schwar\-zen\-berg-Czerny for providing software
  used in this study.

  This work has been supported by the National Science Centre, Poland,
  grant MAESTRO 2016/22/A/ST9/00009. The OGLE project has received
  funding from the Polish National Science Centre grant MAESTRO 
  2014/14/A/ST9/00121.}

\end{document}